\documentclass[12pt]{article}

\usepackage{amssymb}
\usepackage{epsf}

\newlength{\dinwidth}
\newlength{\dinmargin}
\def\be{\begin{equation}}
\def\ee{\end{equation}}
\def\bea{\begin{eqnarray}}
\def\eea{\end{eqnarray}}
\input epsf
\epsfverbosetrue

\begin{document}
\begin{titlepage}
\begin{flushright}
CERN-TH-2003/243\\
\date \\
hep-th/0311244\\
\end{flushright}
\bigskip
\begin{center}
{\bf \Large Comments on Noncommutative Field
Theories}\footnote{Lectures presented at the 9th Adriatic Meeting in
Dubrovnik (Croatia), September 2003 (L.A.-G.) and, in a shortened
form, at String Phenomenology 2003 in Durham (U.K.) July
2003 (M.A.V.-M.).}
\vskip 0.9truecm

Luis \'Alvarez-Gaum\'e
and Miguel A. V\'azquez-Mozo\footnote{Address after January 1st, 2004: 
F{\'\i}sica Te\'orica, Universidad de Salamanca, Plaza de la Merced s/n, E-37008 Salamanca,
Spain.} 

\vspace{1pc}

{\em Theory Division CERN \\CH-1211 Geneva 23\\ Switzerland\\}
{\tt Luis.Alvarez-Gaume@cern.ch, Miguel.Vazquez-Mozo@cern.ch  }

\vspace{5pc}

{\large \bf Abstract}
\end{center}

We discuss some aspects of noncommutative quantum field theories obtained
from the Seiberg-Witten limit of string theories in the presence of an external
$B$-field. General properties of these theories are studied as well as the
phenomenological potential of noncommutative QED.

\vfill

\end{titlepage}

\def\theequation{\thesection.\arabic{equation}}

\section{Introduction}
\setcounter{equation}{0}

Since its formulation by Alain Connes, noncommutative geometry (NCG)
has become a very active and interesting branch of
Mathematics \cite{connesetal}.  In Physics, NCG has had an early impact
in a number of subjects including condensed matter physics
\cite{condmatter} and high energy physics \cite{conneslott}.  In
String Theory, the use of NCG was pioneered by its application by
Witten to string field theory \cite{wittensft}. More recently,
compactifications of string and M-theory on noncommutative tori were
studied in \cite{connesdouglas}. Although quantum field theories in
noncommutative spaces had been the subject of attention \cite{ncqft}, 
a renewed interest in the subject came after the
realization by Seiberg and Witten \cite{seibergwitten} that a certain class
of field theories on noncommutative Minkowski space can be obtained as
particular low-energy limits in the presence of a constant NS-NS
$B$-field.  

Unlike the standard low-energy limit of string theory, the
Seiberg-Witten limit leads to a nonlocal effective theory, where the
interaction vertices are constructed in terms of the nonlocal Moyal
product (see \cite{rev} for comprehensive reviews). In physical terms,
this nonlocality is due to the extended nature of the low energy
excitations, which in fact are rigid rodes whose size depends on the
momentum of the state \cite{dipoles}. It is therefore interesting,
from the field theoretic point of view, to understand how our ordinary
view of field theory changes by the introduction of this particular
type of nonlocality.  Many standard notions and results require
revision, like renormalizability, unitarity, discrete and space-time
symmetries, etc.  Nonetheless, since these theories are obtained from
String Theory, one would expect them to be better behaved than other
kinds of nonlocal theories.

One of the more remarkable results in the subject was obtained by
Minwalla, van Raamsdonk and Seiberg \cite{uv/ir}.  These authors
realized that quantum theories on noncommutative spaces are afflicted
from an endemic mixing of ultraviolet (UV) and infrared (IR) divergences.  Even
in massive theories the existence of UV divergences induce
IR problems, and this leads to a breakdown of the Wilsonian
approach to field theory.  Contrary to some expectations
\cite{snyder}, noncommutativity does not provide a full regularization
of UV divergences, but only of a subsector of the Feynman
graphs. Hence the issue of renormalizability of NCQFT become rather
subtle \cite{renorm}.

In ordinary Quantum Field Theory there are a number of properties
that can be derived from general principles collectively called 
Wightman axioms \cite{wightmanax}.  Among them we can cite
the CPT theorem, the connection between spin-statistics and the cluster
decomposition. The extension of some of these properties to NCQFTs 
is not straightforward \cite{chaichianCPT,cptpapers,ourpaper} and therefore
it would be interesting to study whether this kind of nonlocal
field theories admit an axiomatic formulation in order to gain a better
insight about the extension to NCQFTs of properties like 
the CPT and spin-statistics theorems \cite{ourpaper}.

In this lecture we would like to make a number of remarks on
noncommutative field theories, in particular those obtained from
String Theory through the Seiberg-Witten limit. We will pay special
attention to the analysis of the phenomenological viability of this
kind of field theories. For that we will focus on noncommutative QED
(NCQED). The Standard Model contains Maxwell's theory at low energies
and thus the ``usual'' photon should be recovered in any
noncommutative generalization of QED, independently of how the
Standard Model is embedded into its noncommutative extension.
Among the properties of the QED photon, we will look at its masslessness,
and the fact that the speed of light is constant, i.e. independent of the magnitude
and direction of the photon momentum \cite{speedoflight}.  As we will see,
it is remarkably difficult to obtain that ordinary electromagnetism is
embedded as the low-energy limit of a noncommutative $U(1)$-theory.
In particular, due to UV/IR mixing, it is rather common to
obtain that one of the photon polarizations remains massless while the other
becomes either massive or tachyonic.  In ordinary gauge theories vector
bosons get masses through the Higgs mechanism.  Here the nonlocality of
the interaction terms may lead to a massive photon polarization.  

In order to give sense to NCQED we define it in terms of a softly
broken $\mathcal{N}=4$ noncommutative U(1) gauge theory.  This
provides a construction that makes sense in the UV and IR, and where
we can have control on the UV/IR mixing.  Here we find that unless
some conditions are satisfied by the soft breaking terms, one of the
components of the photon becomes tachyonic.  Even when this disaster
is avoided, one generically gets a completely unacceptable value for
the photon mass, unless one is willing to engage in massive
fine-tuning. We will follow the presentation in our paper 
\cite{ourpaper}, where a more complete list of references is provided.

Before we proceed we would like to clearly state our point of view.
As mentioned above, we will focus here on the type of noncommutative
theories that are obtained from string theory via the Seiberg-Witten
limit.  There are of course other approaches to the problem, and we
would like to briefly make a comparison.  If one follows the
quantization procedure proposed in Refs. \cite{bahnsetal,siboldetal}
the results should be the same, because both approaches agree in the
case of space-space noncommutativity.  Regarding the approach of Ref.
\cite{wessetal}, they extend the Seiberg-Witten map to arbitrary groups, and
their actions are obtained order by order in an expansion in powers of
$\theta$.  Hence if we truncate at a given order, we find the standard
commutative Lagrangian, and a collection of corrections corresponding
the higher dimension operators. This theory is technically
nonrenormalizable and one should not find UV/IR mixing, which occur
only after one has summed to all order in $\theta$, in which case we
would expect to obtain the same results because the Feynman rules are
the same. Other approaches has been studied in \cite{grosseetal}. 

We follow here the ``orthodox" string approach, namely we use the
Feynman rules that follow from String Theory after we take the
Seiberg-Witten limit, in particular we restrict our considerations
always to space-space noncommutativity.  Since the vertices and
Feynman integrands are only modified by sine and cosine functions, the
naive degree of divergence of the theory will not change, and one
should expect some sort of renormalizability to hold once the UV/IR
problems are tamed.  It is possible to extend the Seiberg-Witten limit
to have time-space noncommutativity, but this does not lead to a field
theory but a theory of noncommutative open string \cite{NCOS}.

In the next section we give a short overview of some well-known facts
about NCQFTs, in particular the UV/IR mixing characteristic of these
theories. In Section \ref{CPT} an extension of axiomatic formulation
to NCQFTs is briefly discussed as well as the validity of the CPT
theorem in this type of theories. Section \ref{pi} reviews the IR
problems of NCQED and in Section \ref{final} we study the construction
of such a theory from its softly broken $\mathcal{N}=4$ supersymmetric
extension and the possibility of eliminating tachyonic states in the
spectrum. This section concludes with a discussion of the
phenomenological prospects of NCQED.

\section{Seiberg-Witten limit, dipoles and UV/IR mixing}
\setcounter{equation}{0}
\label{secUVIR}

In \cite{seibergwitten} it was shown 
how noncommutative field theories are
obtained as a particular low-energy limit of open 
string theory on D-brane backgrounds in the presence of
constant NS-NS $B$-field. In this case, the endpoints
of the open strings behave as electric charges in the
presence of an external magnetic field $B_{\mu\nu}$
resulting in a polarization of the open strings. Labeling by
$i=1,\ldots,p$  the D-brane directions and assuming $B_{0i}=0$,
the difference between the zero modes of the string
endpoints is given by \cite{dipoles}
\begin{eqnarray}
\Delta X^{i}=X^{i}(\tau,0)-X^{i}(\tau,\pi)=(2\pi\alpha')^2 
g^{ij}B_{jk}p^{k},
\label{sw}
\end{eqnarray}
where $g_{\mu\nu}$ is the closed string or $\sigma$-model metric and
$p^{\mu}$ is the momentum of the string.
In the ordinary low-energy limit, where $\alpha'\rightarrow 0$ while
$g_{\mu\nu}$ and $B_{\mu\nu}$ remain fixed, the distance $|\Delta X|$
goes to zero and the effective dynamics is described by a theory of
particles, i.e. by a commutative quantum field
theory.

There are, however, other possibilities of decoupling the massive
modes without collapsing at the same time the length of the
open strings. Seiberg and Witten proposed to consider a low-energy
limit $\alpha'\rightarrow 0$ where both $B_{ij}$ and
the open string metric
\begin{eqnarray}
G_{ij}=-(2\pi\alpha')^2 (Bg^{-1}B)_{ij}
\end{eqnarray}
are kept fixed.
Introducing the notation $\theta^{ij}=(B^{-1})^{ij}$, the separation
between the string endpoints  can be expressed as:
\begin{eqnarray}
\Delta X^{i}=\theta^{ij}G_{jk}p^{k},
\end{eqnarray}
fixed in the low-energy limit. The resulting low-energy
effective theory is a noncommutative field theory with 
noncommutative parameter $\theta^{ij}$.  In physical terms
the Seiberg-Witten limit corresponds to making the string
tension go to infinity, while and balancing it with the
Lorentz force on the string-ends caused by the external
magnetic field.  This limit makes the string rigid and
with a finite length that depends on its momentum.

The previous analysis was confined to
situations in which the $B_{0i}$ components are set to zero. 
The result is a noncommutative
field theory with only space-space noncommutativity. From a purely
field-theoretical point of view it is possible
to consider also noncommutative theories where the time coordinate
does not commute with the spatial ones, i.e.
$\theta^{i0}\neq 0$. In this case, however, non-locality
is accompanied by a breakdown of unitarity reflected
in the fact that the optical theorem is not satisfied
\cite{gomismehen,AGBZ}. In addition there is no well-defined
Hamiltonian formalism (see, however, the alternative approaches of 
\cite{bahnsetal,siboldetal}). From the string theory point of view,
taking the Seiberg-Witten limit with $B_{0i}\neq 0$ results in a lack
of decoupling of closed string modes. In the resulting low energy
noncommutative field theory the violation of the optical theorem can
be formally solved by including the undecoupled string modes that,
however, have negative norm \cite{AGBZ}. In the following we will 
restrict our attention to the case of space-space noncommutativity,
$\theta^{0i}=0$.

In the Seiberg-Witten limit we obtain therefore field theories
on a quantum plane, the coordinates $x^{\mu}$ do not commute
but rather satisfy:
\bea
\label{commrel}
[x^{\mu},x^{\nu}]&=& i \theta^{\mu\nu},
\eea
with
\be
\label{theta}
\theta^{\mu\nu}= \pmatrix{0 & 0 & 0 & 0 \cr 
0 & 0 & \theta & 0 \cr              
0 & -\theta & 0 & 0 \cr 
0 & 0 & 0 & 0 }.
\ee
The action for these field theories looks the same as for commutative
theories except that functions are multiplied in terms of the Moyal
product:
\be
\label{moyal}
f(x)\star g(x)=f(x)e^{{i\over 2}\theta^{\mu\nu}\stackrel{\leftarrow}{\partial}_{\mu}
\stackrel{\rightarrow}{\partial}_{\nu}}g(x).
\ee
Using the Fourier transform of (\ref{moyal}) we can write down the
Feynman rules for a scalar field theory containing a $\varphi^n_\star$-vertex.
The result is:
\be
\label{vertex}
\int d^{d}x\, \varphi(x)^n_\star =\int {d^{d}k_1\over (2\pi)^{d}} 
\ldots {d^{d}k_n\over(2\pi)^d}\,(2\pi)^d \delta\left(\sum_{j=1}^{n}k_{j}\right)\,
\tilde{\varphi}(k_1)\ldots\tilde{\varphi}(k_n) 
e^{-{i\over 2}\sum_{i<j} k_i\cdot \tilde{k}_j},
\ee
with $\tilde{k}^{\mu}\equiv \theta^{\mu\nu}k_{\nu}$ and
$\tilde{\varphi}(k)$ the Fourier transform of $\varphi(x)$.  

\begin{figure}[h]
\centerline{ \epsfxsize=4.0truein \epsffile{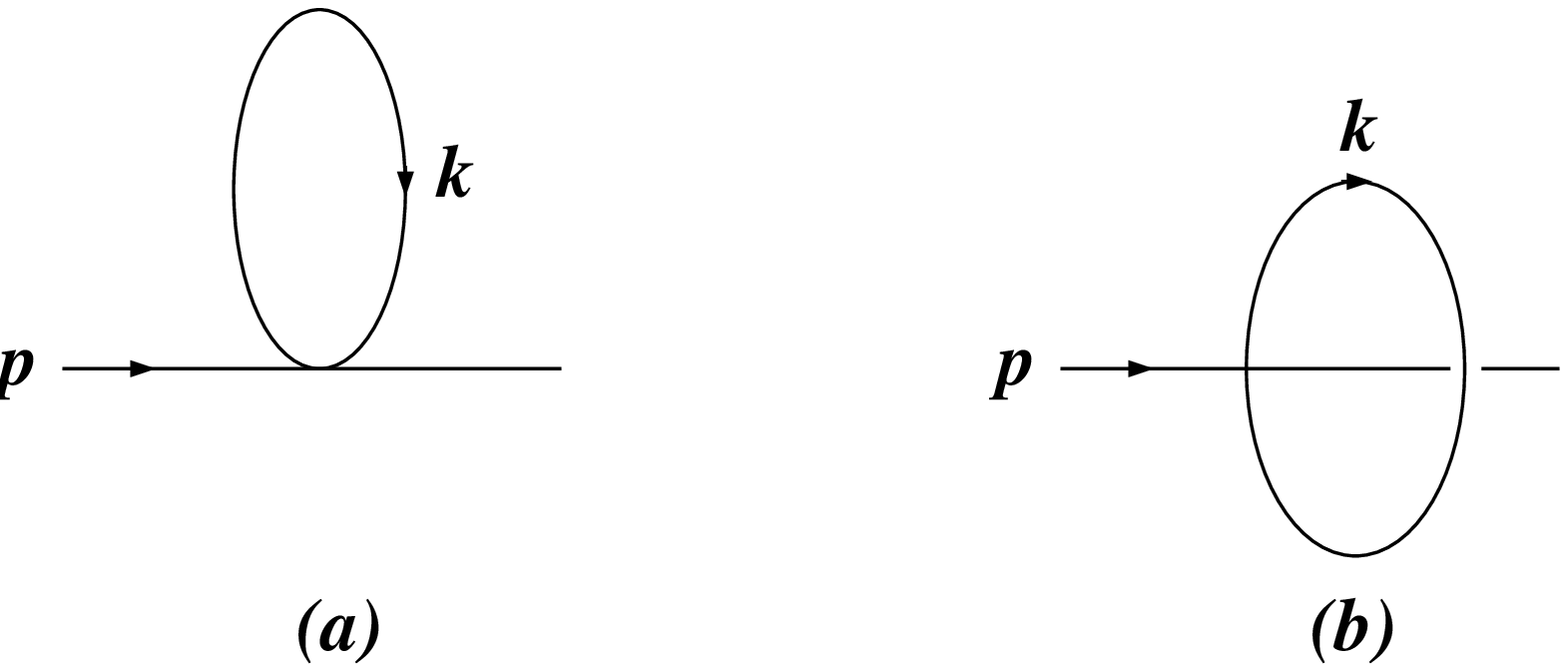}}
\caption{Planar (a) and nonplanar (b) contributions to the mass renormalization 
in $\lambda\varphi_{\star}^{4}$.} 
\label{fr}
\end{figure}

The phases in (\ref{vertex}) are at the origin of the UV/IR mixing
\cite{uv/ir}.  Because of these phases depend on the incoming momenta
in the vertex, planar and nonplanar Feynman diagrams will have different
contributions. As a matter of
example one can consider the mass renormalization in $\lambda
\varphi^4_\star$-theory. Whereas in the commutative theory there is a
single diagram contributing to one loop, the new Feynman rules produce
two contributions (see Fig. 1). One of them is the quadratically divergent planar
diagram which is identical to the commutative one except for a different 
combinatorial factor. Together with this there is the ``non-planar"
contribution (see \cite{uv/ir} for details) which has the form:
\be
\label{nonplanar}
\Pi(p)_{\rm nonplanar}={\lambda\over 6}
\int {d^4k \over (2\pi)^4}{e^{ik.{\tilde p}}\over k^2+m^2}.
\ee
As long as the external momentum $p$, or rather $\tilde p$, is
nonvanishing, the integral converges due to the rapidly oscillating
phase at large loop momentum. Exponentiating the denominator in
(\ref{nonplanar}) using a Schwinger parameter and introducing a
Schwinger cutoff $\Lambda$ we find that the nonplanar diagram has an
effective momentum-dependent cut-off given by
\begin{eqnarray} 
{1\over \Lambda_{\rm eff}^2}={1\over
\Lambda^2} +{\tilde p}^2, 
\end{eqnarray}
This clearly shows how UV divergences of (\ref{nonplanar}) are
transformed into IR ones. After the UV cutoff $\Lambda$ is sent to
infinity, the quadratic divergence will reappear in the IR limit
$\tilde{p}\rightarrow 0$.  At fixed cutoff, on the other hand, the
UV/IR mixing reflects in that the two limits $\Lambda\rightarrow
\infty$ and $\theta\rightarrow 0$ do not commute. 
It is thus clear that in general we will have problems defining
low-energy Wilsonian effective actions since UV and IR scales
do not decouple.

The phenomenon of UV/IR mixing has some resemblance with Quantum
Gravity or String Theory, and is probably one of the reasons why NCQFTs
have received so much attention.  When we consider General Relativity,
an object with a given energy $E$ has two lengths associated with it:
one is the Compton wavelength as in ordinary field theory
$\hbar/E$. At the same time it also has its Schwarzschild radius $G_N
E$. Obviously as the energy grows there is a point where the
Schwarzschild radius becomes bigger than the Compton wavelength, and
certainly at this point our standard notions of quantum field theory
no longer apply.  

If we consider the origin of the UV/IR mixing, the analogy is very
appealing.  In the loops of the NCQFTs we will have particles of very
high energy and thus very short Compton wavelength. Since the
fundamental objects in the theory are dipoles, the states running in
loops have also an associated length of order $\theta p$, growing with
the energy.  If we define the theory with a sharp momentum space
cutoff $\Lambda$, the longest dipole has size $\Lambda\theta$. This
rod-like structure of the noncommutative quanta breaks Lorentz
invariance, as it is already clear from the commutation relations
(\ref{commrel}) with the noncommutativity parameter given by
(\ref{theta}). However if we consider momenta $0\le p \lesssim
1/(\Lambda\theta)$ Lorentz invariance and the commutative theory
should be recovered, since these momenta correspond to length scales
bigger than the size of the largest dipole and therefore one cannot
probe the ``dipolar'' structure of the excitations of the
noncommutative theory.

\section{Residual symmetries, general properties and  discrete symmetries}
\setcounter{equation}{0}
\label{CPT}

Looking at the commutation relations (\ref{commrel}) we see that NCQFTs
are invariant under translations and therefore states can still be
labeled by their four-momentum eigenvalues. On the other
hand\footnote{ In the general case, $\theta^{\mu\nu}$ is determined by two
vectors, the electric  and magnetic  components $\theta^{0i}$ and
$\epsilon^{ijk}\theta^{jk}$.  If they are not
collinear, the Lorentz group is completely broken.}, given the form of
$\theta$ in Eq. (\ref{theta}), the Lorentz group is broken from O(1,3)
to O(1,1)$\times$SO(2).

For generic theories with these reduced symmetry one should not expect
a connection between spin and statistics, and it should not be hard to
construct theories with exotic statistics. It is thus an interesting
question whether basic theorems like CPT which hold in local
relativistic field theories will continue to hold in this context.
Since we consider these theories as obtained from the Seiberg-Witten
limit, it is reasonable to ask this question within String Theory.
The CPT theorem in string theory has been investigated by several
groups \cite{CPTstring}.  If the parent string theory satisfies the
CPT theorem in perturbation theory, and since the constant background
$B$-field is CPT-even, it is reasonable to expect that the
noncommutative quantum field theory obtained in the Seiberg-Witten
limit should also preserve CPT.  

At the level of the noncommutative field theory it is also expected to
have CPT-invariance for theories with $\theta^{0i}=0$. As mentioned
above, these theories preserve perturbative unitarity.  In ordinary
quantum field theory there is an intimate connection between unitarity
and CPT-invariance. Indeed, if the condition of asymptotic
completeness holds, it can be shown \cite{yf,hr} that the $S$-matrix
can be written in terms of the CPT operator of the complete theory
$\Theta$ and the corresponding one for the asymptotic theory
$\Theta_{0}$:
\begin{eqnarray}
S=\Theta^{-1}\Theta_{0}.
\end{eqnarray}
The unitarity of the $S$-matrix follows then from the antiunitarity of
$\Theta$ and $\Theta_{0}$. A theory with CPT invariance is therefore
likely to be unitary.

There are several proofs of the CPT theorem for ordinary QFTs.  We
find however that the deeper and more elegant one is due to Jost
\cite{jost,wightmanax}.  Few ingredients are required. One only needs
the theory to satisfy the Wightman axioms, essentially Poincare invariance,
uniqueness of the vacuum, positivity of the energy and microscopic
causality.  With these conditions it is shown that
the Wightman functions\footnote{Wightman functions are vacuum expectation values of products
of fields without time-ordering, namely 
\begin{eqnarray*}
W_{n}(x_{1},\ldots,x_{n})=\langle\Omega|\Phi(x_{1})\ldots\Phi(x_{n})|\Omega\rangle
\end{eqnarray*}
where $|\Omega\rangle$ is the true vacuum of the theory.} admit an
analytic continuation that is invariant under the complexified Lorentz
group.  The standard Lorentz group has four sheets, one connected to
the identity, and the other three obtained by applying to it P,
T and PT.  However, the complexified
Lorentz group contains only two sheets, and the one
obtained by applying the full space-time inversion (PT) is connected
with the identity.  Expressing invariance under this transformation
essentially amounts to the proof of the CPT theorem.  

In spite of the reduced space-time symmetry, this proof can be
extended to NCQFTs, at least those with space-space
noncommutativity\footnote{A different proof that applies also the the
time-space noncommutativity can be found in \cite{chaichianCPT}.}.
The key ingredient lies in the fact that for NCQFT microcausality is
defined only with respect to the O(1,1) factor of the space-time
symmetry group. Given the fact that this group has a structure very
similar to the full O(1,3) Lorentz group, Jost's proof is carried out
to the noncommutative case without any problem (see \cite{ourpaper}
for the details).

Hence, although in general one should expect serious problems with
nonlocal theories, for the type of nonlocality produced by the
Seiberg-Witten limit we nevertheless recover the standard form of the
CPT theorem.  The spin-statistics connection is more subtle. If the
type of representations of the reduced Lorentz group in NCQFTs are
obtained as reductions from standard representations, it is likely
that the same construction goes through.  However, if we start with
other representations, not inherited from higher dimensions, exotic
statistics may easily occur.

\section{The infrared problems of noncommutative QED}
\setcounter{equation}{0}
\label{pi}

As we discussed in Section \ref{secUVIR}, much of the interesting
physics of NCQFTs comes from UV/IR mixing. This lack of decoupling of
the different scales in the theory might pose a serious problem to
phenomenology, since noncommutative effects can show up at low
energies interfering with standard model predictions.

Since, apart from gravity, electromagnetism is the only long range
interaction at low energies, it seems that QED would be the perfect
test bench for the phenomenology of NCQFT. The noncommutative version of QED
can be easily constructed by deforming ordinary QED with the introduction
of Moyal products. Here we will consider the simplest case of pure
NCQED with action
\begin{eqnarray}
S_{\rm NCQED}={1\over 4g^2}\int d^{4}x\,F_{\mu\nu}F^{\mu\nu}
\end{eqnarray}
where $g$ is the coupling constant and
\begin{eqnarray}
F_{\mu\nu}=\partial_{\mu}A_{\nu}-\partial_{\nu}A_{\mu}-i(A_{\mu}\star
A_{\nu}-A_{\nu}\star A_{\mu})
\end{eqnarray}
Because of the quadratic $\theta$-dependent term in the definition of
$F_{\mu\nu}$, the noncommutative photon self-interacts unlike
the case of ordinary QED. The corresponding Feynman rules are very similar
to those of nonabelian Yang-Mills where the group structure constants
are replaced by trigonometric functions of the incoming momenta (see Fig. 2).

\begin{figure}[h]
\centerline{ \epsfxsize=5.5truein \epsffile{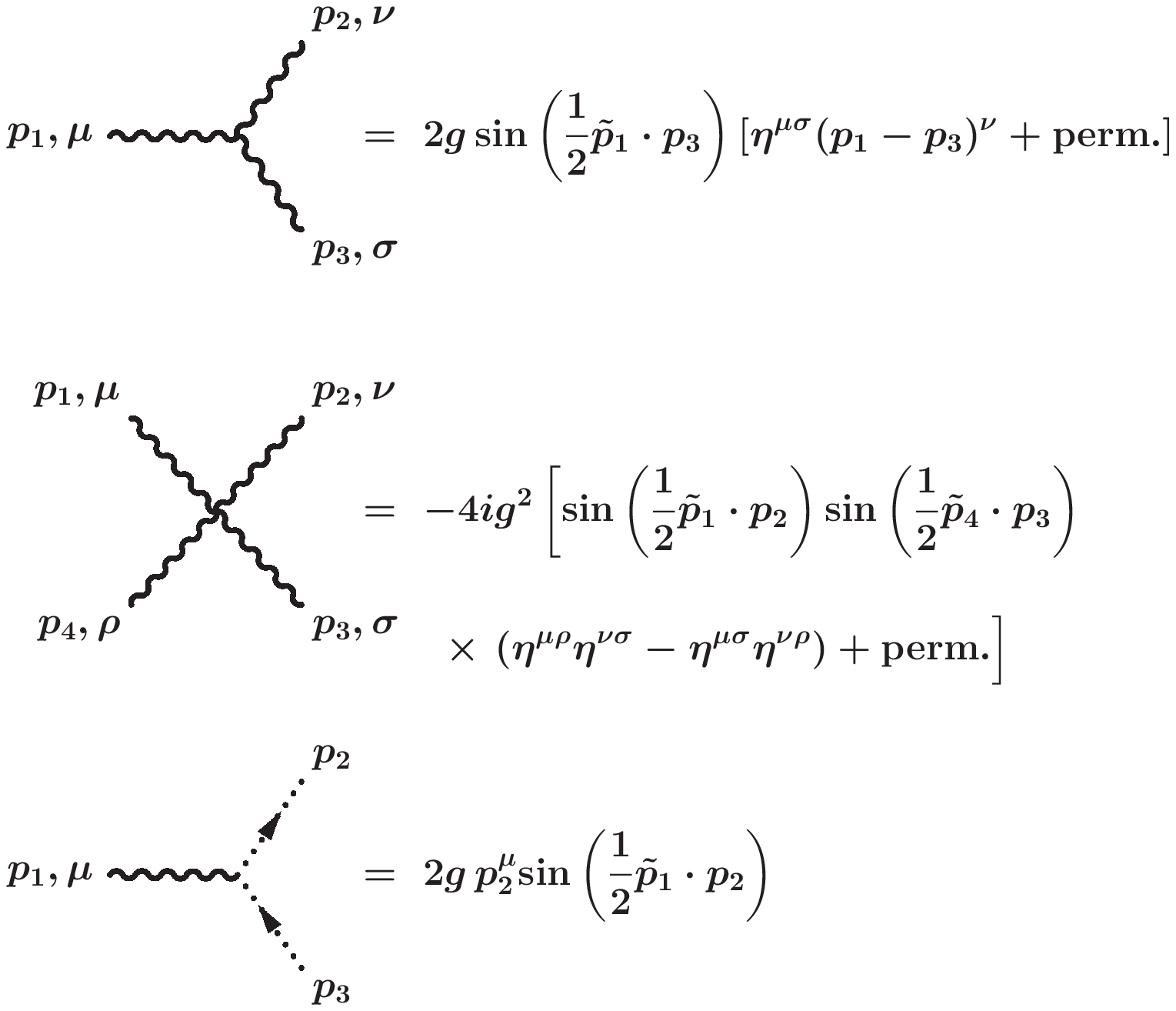}}
\caption{Interaction vertices for NCQED. The wavy line represents the photon and
the line of points the Fadeev-Popov ghost.} 
\end{figure}

At tree level, noncommutative corrections to ordinary QED can be made
small and compatible with experimental bounds as long as the
noncommutative energy scale $1/\sqrt{\theta}$ is chosen large
enough. This is due to the fact that noncommutativity only appears in
the form of global $\theta$-dependent phases which have a smooth
commutative limit. The situation is radically different once one-loop
effects are taken into account, due to UV/IR mixing. The kind of
problems encountered show up in the calculation of the one-loop
corrected dispersion relation for the photon in NCQED which 
results to be \cite{disperrel}
\begin{eqnarray}
\omega(\vec{p}\,)^2=\vec{p}^{\,2}-{2g^2\over \pi^2}{1\over p\circ p},
\label{dis1}
\end{eqnarray}
where $p\circ p\equiv \theta^2(p_1^2+p_2^2)$.  The divergence at low
transverse momentum in (\ref{dis1}) is the result of a UV quadratic
divergence that reappears in the IR as a result of UV/IR mixing.  It
seems therefore that the disastrous situation implied by the previous
dispersion relation could be overcome by completing noncommutative QED
with another theory softer in the UV.

This was attempted by embedding NCQED into $\mathcal{N}=1$
noncommutative supersymmetric QED and breaking supersymmetry softly
at a scale $M$ by adding a mass term to the gaugino. The resulting 
dispersion relation, however, eliminates the IR divergence although it
leaves behind a finite tachyonic mass for the photon \cite{cetal}
\begin{eqnarray}
\omega(\vec{p}\,)^2 \approx \vec{p}^{\,2}-{g^2 M^2\over 2\pi^2}.
\end{eqnarray}
In spite of having removed the divergence, the situation has not improved
at all, since the tachyonic mass of the photon only depends on the 
soft-breaking mass of the gaugino and not on the noncommutative parameter 
$\theta$. Therefore it cannot be made small by taking the noncommutative 
energy scale large. On the contrary if, for phenomenological reasons, we 
set $M\sim 1\,\mbox{TeV}$ we find a tachyonic photon with a mass much outside 
the current experimental bounds.

As we already mentioned, the source of all the trouble with NCQED is
the mixing of UV and IR scales, which induces IR singularities in
nonplanar amplitudes. One possible way to tame this problem is by
defining the noncommutative theory with an sharp UV cutoff
$\Lambda>1/\sqrt{\theta}$. In this case, the noncommutative scale
gets, in a sense, ``corrected'' due to one loop effects and
noncommutative effects start being relevant already at scales of order
$1/(\theta\Lambda)$. Moreover, because of the regularization of the UV
singularities provided by the cutoff, the commutative theory is
recovered at scales $E\ll 1/(\theta\Lambda)$, including its Lorentz
invariance.

Therefore, it seems that this might provide a way to define NCQED at
low energies avoiding the problems of the emergence of
tachyons. Unfortunately \cite{ourpaper}, there are additional
difficulties associated with this regularization scheme. In
particular, apart from the lattice, a sharp cutoff $\Lambda$ of the
type required (either a cutoff in momenta or a Schwinger cutoff) leads
to violations of gauge invariance. This can only be avoided by
considering ``mixed'' cutoffs which combines a sharp cutoff with
dimensional regularization.  In the case of NCQED this scheme works
fine for the one-loop polarization tensor where gauge invariance is
preserved and the result for ordinary QED recovered at low
momentum \cite{ourpaper}. Nevertheless, its extension to other
amplitudes or higher loops is more problematic.

\section{Some phenomenological considerations on NCQED}
\setcounter{equation}{0}
\label{final}

A second alternative, that we will pursue here, is to ameliorate the
IR problems of NCQED by looking for high energy completion of the
theory which would be free of UV divergences. In particular, let us
consider $\mathcal{N}=4$ U(1) noncommutative super-Yang-Mills, which
is believed to be finite \cite{jones} as its
commutative counterpart. In this case, instead of a single U(1) gauge
vector field we have one $\mathcal{N}=1$ vector multiplet together
with three scalar multiplets in the adjoint representation. NCQED can
be then recovered at low energies by breaking supersymmetry softly by
adding masses $M_f$ to the gauginos and $M_s$ to the
scalars \cite{ssb}. This provides a construction that makes sense in
the UV and IR, and where we can have control on the UV/IR mixing.

With this setup, we can proceed to compute the one-loop polarization
tensor for the photon $\Pi_{\mu\nu}(p)$. We will work in Euclidean
signature and rotate back to Minkowski at the end of the
calculation. On symmetry grounds it has the form
\begin{eqnarray}
\Pi_{\mu\nu}(p)=\Pi_{1}(p)\left(p^2\delta_{\mu\nu}-p_{\mu}p_{\nu}\right)
+\Pi_2(p){\tilde{p}_{\mu}\tilde{p}_{\nu}\over \tilde{p}^2}
\label{polarization}
\end{eqnarray}
where $\tilde{p}^{\mu}=\theta^{\mu\nu}p_{\nu}$. It is important to
notice that, due to the antisymmetry of $\theta^{\mu\nu}$, the extra
piece on the right-hand side in Eq. (\ref{polarization}) is transverse and the
Ward identity $p^\mu\Pi_{\mu\nu}(p)=0$ is satisfied.

Now we can proceed to compute the functions $\Pi_1(p)$ and $\Pi_2(p)$
at one loop for the theory with soft-breaking mass terms. Using the
background field method \cite{kt} and working in dimensional
regularization in the $\overline{\mbox{MS}}$ scheme the results are
\begin{eqnarray}
\Pi_1(p)&=& {1\over 4\pi^2}\int_{0}^{1}dx\left\{\left[4-(1-2x)^2\right]\left[
{1\over 2}\log\left({\Delta_v\over
4\pi\mu^2}\right)+K_{0}\left(\sqrt{\Delta_v}|\tilde{p}|\right)\right]
\right. \nonumber \\
&-&\left[1-(1-2x)^2\right]\sum_{f}\left[ {1\over
2}\log\left({\Delta_f\over
4\pi\mu^2}\right)+K_{0}\left(\sqrt{\Delta_f}|\tilde{p}|\right)\right]
\nonumber \\
&-&\left.{1\over 2}(1-2x)^2\sum_{s}\left[ {1\over
2}\log\left({\Delta_s\over
4\pi\mu^2}\right)+K_{0}\left(\sqrt{\Delta_s}|\tilde{p}|\right)\right]
\right\},
\label{pi1}
\end{eqnarray}
and
\begin{eqnarray}
\Pi_2(p)&=& -{1\over \pi^2}\int_{0}^{1}dx\left[\Delta_v K_2 
\left(\sqrt{\Delta_v}|\tilde{p}|\right)-
\sum_{f}\Delta_f K_2\left(\sqrt{\Delta_f}|\tilde{p}|\right)\right. \nonumber \\
&+&\left. {1\over 2}\sum_{s}\Delta_s
K_2\left(\sqrt{\Delta_s}|\tilde{p}|\right) \right].
\label{pi2}
\end{eqnarray}
Here $\mu$ is the dimensional regularization energy scale and the
subindices $v$, $f$ and $s$ indicate respectively the contributions
from the vector-ghost system, fermions and scalars. In addition we have defined
\begin{eqnarray}
\Delta_v&=&x(1-x)p^2,\nonumber \\
\Delta_f&=&M_f^2+x(1-x)p^2, \nonumber \\
\Delta_s&=&M_s^2+x(1-x)p^2,
\end{eqnarray}
with $M_f$, $M_s$ the soft-breaking masses.

With Eqs. (\ref{pi1}) and (\ref{pi2}) we can easily compute the
dispersion relation by looking at the poles of the full propagator
$G_{\mu\nu}(p)$, once the one loop 1PI parts are resumed:
\begin{eqnarray}
G_{\mu\nu}(p)&=&{ig^2\over p^2}\left[\mathbf{1}+{-g^2\over p^2}\mathbf{\Pi}(p)+
\left({-g^2\over p^2}\right)^2\mathbf{\Pi}(p)^2+\ldots\right]_{\mu\nu}
\end{eqnarray}
where $\mathbf{1}$ is the $4\times 4$ identity matrix and
$\mathbf{\Pi}(p)$ is a matrix notation for the polarization tensor
in Eq. ({\ref{polarization}). After a straightforward calculation we find
\begin{eqnarray}
G_{\mu\nu}(p)&=&{ig^2p_\mu p_\nu\over p^2}+
{ig^2\over p^2\left[1+g^2\Pi_1(p)\right]}\left(\delta_{\mu\nu}-{p_{\mu}p_{\nu}\over p^2}
\right) 
\label{resumed}
\\
&+&\left\{{ig^2\over p^2\left[1+g^2\Pi_1(p)\right]+g^2\Pi_2(p)}-
{ig^2\over p^2\left[1+g^2\Pi_1(p)\right]}\right\}{\tilde{p}_{\mu}\tilde{p}_{\nu}\over 
\tilde{p}^2}.\nonumber 
\end{eqnarray}

Unlike the case of ordinary QED in Eq. (\ref{resumed}) we have two
sources of poles in the full photon propagator. On the one hand we
find the usual solution
\begin{eqnarray}
 p^2\left[1+g^2\Pi_1(p)\right]=0,
\end{eqnarray}
which gives rise to the usual massless dispersion relation for the
photon, $p^2=0$.  Together with this we also find a second pole
associated with photon polarizations along the vector $\tilde{p}^\mu$:
\begin{eqnarray}
p^2\left[1+g^2\Pi_1(p)\right]+g^2\Pi_2(p)=0.
\end{eqnarray}
In order to extract the dispersion relation we perform the rotation
back to Minkowski signature by replacing $p^2\rightarrow -p^2$ and
$\tilde{p}^2\rightarrow p\circ p$. Using the low momentum expansion of
Eqs. (\ref{pi1}) and (\ref{pi2}) we find the dispersion relation for the 
polarizations along $\tilde{p}^\mu$ for low momentum
\begin{eqnarray}
\omega(\vec{p})^2\approx \vec{p}^{\,2}-{g^2\over 2\pi^2}\left(
\sum_{f}M_{f}^2-{1\over 2}\sum_{s}M_s^2\right).
\label{disrel4}
\end{eqnarray}

Unlike the case in which NCQED is completed in the UV by
$\mathcal{N}=1$ U(1) noncommutative super-Yang-Mills \cite{cetal}, here
we can avoid a tachyonic photon by appropriately tuning the soft
breaking masses in Eq. (\ref{disrel4}), i.e. by demanding
\begin{eqnarray}
\sum_{f}M_{f}^2-{1\over 2}\sum_{s}M_s^2\leq 0.
\label{ineq}
\end{eqnarray}
Unfortunately, a tuning of this quantity to zero does not result in a
massless photon polarization, as we would like to recover at low
energies. When the inequality (\ref{ineq}) is saturated the leading
term in the expansion of $\Pi_2(p)$ around $\vec{p}=0$ is negative,
and we find a dispersion relation with negative energy squared for low
momentum photons.  Therefore one is forced to a finite value of the
quantity on the left hand side of Eq. (\ref{ineq}), i.e. to a massive
photon polarization. Using the current bounds for the photon
mass \cite{pdb}, one has to engage in a massive fine tuning of the
soft breaking masses
\begin{eqnarray}
{1\over 2}\sum_{s}M_s^2-\sum_{f}M_{f}^2\lesssim 10^{-32}\,\,\mbox{eV}^2.
\end{eqnarray}

This result is not affected by the addition of matter in the
fundamental representation of U(1). In the calculation of the one-loop
polarization tensor fundamental fields in the loop only contribute to
planar diagrams. Since the function $\Pi_{2}(p)$ in
Eq. (\ref{polarization}) is solely determined by non-planar
diagrams the only effect of the fundamental fields is in modifying the
running of the coupling constant through the function $\Pi_{1}(p)$.

Even if the problem of a tachyonic photon can be avoided by this
un-natural fine tuning of the mass scales, the dispersion relation of
photons with polarizations along $\tilde{p}^\mu$ will be different
from the standard relation $\omega(\vec{p})=|\vec{p}|$ of photons with
polarizations orthogonal to $\tilde{p}^\mu$. This implies that in the
construction of NCQED we are studying here there is a phenomenon of
birefringence associated with the fact that the dispersion relations
(and therefore the speed of propagation) of photons with different
polarizations are different (cf. \cite{jackiw}).

After our analysis we have to conclude that the phenomenological
perspectives of NCQED look rather poor. In our attempt to eliminate
the tachyonic polarization of the photon we have been lead to
massive photon polarizations and birefringence, at the prize also of 
a huge fine tuning of the masses of the soft breaking masses.

To summarize, here we have studied the problem of making sense out of
NCQED at low energies, as derived from string theory in the
Seiberg-Witten limit. To ameliorate the hard IR problems that afflict
this theory we have completed it in the UV by $\mathcal{N}=4$
noncommutative U(1) super-Yang-Mills, softly broken by mass terms for
the gauginos and scalars.  Our conclusions regarding the
phenomenological viability of such a theory are, however, rather
negative. We found that tachyons can be avoided only by allowing a
massive polarization for the photon. This requires also a tremendous
fine tuning of the soft-breaking masses. It seems, therefore, that any
attempt to extract phenomenology from this theory should be postponed
to find a formulation of the theory that can describe at least the
rough features of the world we live in.

\section*{Acknowledgments}

We thank the organizers of the 9th Adriatic Meeting and the String
Phenomenology 2003 Conference for the opportunity of presenting this
work. We are also thankful to J.L.F. Barb\'on, M. Chaichian,
J.M. Gracia-Bond\'{\i}a, K.E. Kunze, D. L\"ust, R. Stora and J. Wess
for useful discussions. M.A.V.-M. acknowledges support from Spanish
Science Ministry Grant FPA2002-02037.

\end{document}